# Towards a Calculus of Object Programs

**Bertrand Meyer**
**ETH Zurich, ITMO & Eiffel Software**
se.ethz.ch

## Abstract

Verifying properties of object-oriented software requires a method for handling references in a simple and intuitive way, closely related to how O-O programmers reason about their programs. The method presented here, a Calculus of Object Programs, combines four components: *compositional logic*, a framework for describing program semantics and proving program properties; *negative variables* to address the specifics of O-O programming, in particular qualified calls; the *alias calculus*, which determines whether reference expressions can ever have the same value; and the *calculus of object structures*, a specification technique for the structures that arise during the execution of an object-oriented program.

The article illustrates the Calculus by proving the standard algorithm for reversing a linked list.

## 1  Addressing the specifics of object-oriented software

Object-oriented programming predominates today; the verification methods we apply should reflect its distinctive properties. Much of the available work, however, fails to take into account the specifics of the object-oriented approach, in particular the "general relativity" principle which makes every operation dependent on a "current object" known only at run time and potentially different for every execution or evaluation.

The most critical obstacle, for the practice of verification, is that there is still no easily applicable approach to handle the manipulation of *references* (pointers), which plays a central role in the practice of O-O development. Separation logic, the method that has attracted the most attention, rests on an extensive model of the heap, requires extensive program annotations, and fails to take advantage of the abstraction mechanisms that define object technology. A typical example is Bornat's important work on "proving pointer programs" [1], which does not consider any object-oriented mechanisms — in fact not even routine calls, O-O or not — and focuses its discussion on modeling remote field assignments, $x.a := c$, a mechanism that no careful object-oriented programmer would use. (The appropriate idiom, whether or not the language imposes it, is to go through a call to a setter procedure $x.set\_a\,(c)$, or a semantically equivalent variant such as a property setter in C#.)

The present discussion describes an approach to verifying object-oriented programs with particular emphasis on the handling of references as required for linked data structures. The techniques closely follow the way object-oriented programmers think about their programs; it uses standard annotations (contracts) of the form already present in Eiffel, Spec# or JML, with only small extensions to the concepts of axiomatic (Hoare-style) specification. It retains the possibility of evaluating assertions at run time, for testing purposes, in addition to using them for static verification. Although the present paper presents the concepts only, the integration into a state-of-the-art proof seems within reach.

The approach includes four components:

- *Compositional logic* (section 3), which describes the semantics of program elements in terms of their effects on program values, generalizing the assertions of Hoare-Dijkstra semantics to expressions of arbitrary types.

- The notion of *negative variable* (section 4), which provides a simple machinery to model the distinctive properties of object-oriented programming, making it possible in particular to reason on properties of the fundamental operation of object-oriented programming, the call $x.f\,(args)$.

- The *alias calculus*, an automatic approach (not relying on annotations) to determine that two given expressions in a program can never denote the same object. The alias calculus was presented in an earlier paper [16]; section 5 summarizes its results and its application to the present work.

- The *calculus of object structures* (section 6), a set of techniques for describing properties of run-time structures involving references. Reasoning effectively about object structures requires suitably abstract models; the calculus defines these abstractions, in particular through the integral operator $\int$, and the associated semantic rules.

The "Calculus of Object Programs" is the combination of these four techniques. It yields, as an example, a simple proof of a program known to be challenging for verification: linked list reversal. To enable the reader to understand right away how the techniques work, this proof appears in section 2, where each step includes a forward reference to the formal rule that justifies it. Sections 3 to 6 detail these rules; section 7 is a comparison with other approaches, section 8 concludes, and appendix A provides some supplementary theoretical background.

Starting with the example should enable the reader to see the fundamental simplicity of the method, and encourage the study of the theory in the remainder of the paper.

The approach has limitations, detailed in section 8; for example, it does not yet address inheritance. Also, the ideas have not yet been implemented; integrating them into a practical verification environment [31] will, we hope, show their practicability and scalability. Another possible criticism is that not much attention has been devoted so far to modular provability. In spite of these limitations, the Calculus of Object Programs presented here may hold some of the elements of a simple method for verifying programs that routinely manipulate sophisticated object structures.

## 2  A proof: linked list reversal

The example proof addresses an important and typical problem involving somewhat intricate manipulations of references: the in-place reversal of a linked list.

As evidence that many people consider it tricky, we note that it is a staple interview question for programmers; dozens of Web pages, which one will readily find through a search for terms such as "list reversal algorithm", present variants of the solution, for the benefit of job candidates preparing such interviews. (A blog entry [17] discusses some of these pages, noting that they typically fail to mention the loop invariant even though it is the key to understanding the algorithm.)

Although the steps are simple, we will perform the key part of the proof in almost full detail, in the way one would present a proof of Euclid's algorithm in an introductory course on axiomatic semantics. The intent is to show that the techniques presented here allow programmers to reason formally about programs manipulating linked data structures as simply and naturally as about traditional programs involving just integers and booleans.

Terminology note: an object is made of a number of elementary values known as *fields*. Of direct interest for the present discussion are fields of reference types (rather than of basic types such as integers). Every field in an object corresponds to an *attribute* of the associated class. Attributes are also called "member variables" (or "fields", although this term may cause confusion between the static and dynamic views).

## 2.1 Algorithm idea

The goal is to reverse a list of cells (of type *LINKABLE*) linked to each other through fields labeled *right*; the first cell is accessed through the field *first* of the list class:

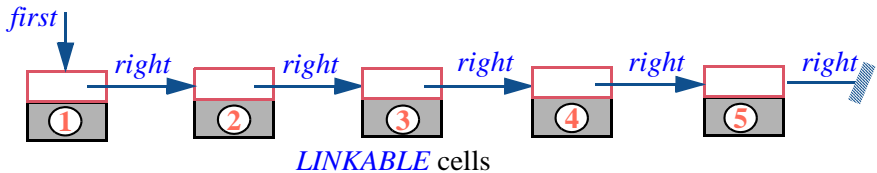

(The figures and part of the discussion are taken from an introductory programming textbook [15].) As illustrated, each cell contains both a *right* link and some other information, shown here as just an integer. We will assume that the structure induced by the *right* links is acyclic; this property, formalized below, must remain invariant throughout the algorithm. The desired final situation is:

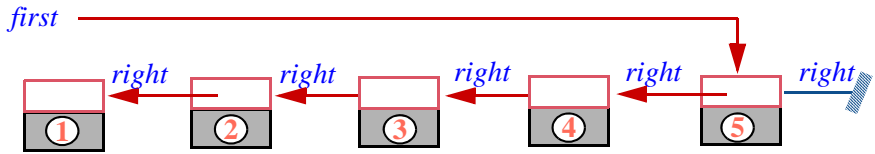

By convention, the algorithm reorders the cells by changing their *right* fields, but does not change the rest of the cell's contents (so that in this example each cell in the final picture is the same object as the one that had the same integer identifier in the original). Other variants are of course possible.

The best way to understand the basic idea of the algorithm, which relies on a loop, is to consider the state of the data structure after a typical iteration of the loop:

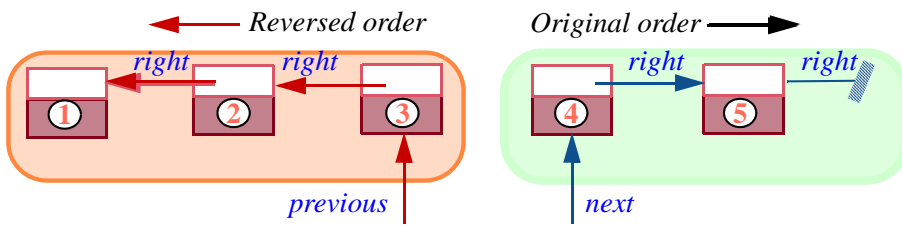

In this intermediate state, we actually have two lists, accessible through the local variables *previous* and *next*. The key property, as illustrated, is that the first list contains an initial subset of the original sequence, but now in reversed order, and the second list contains the remaining elements, in their original order. Then the task of the loop body is to preserve this property but move the boundary between the two lists by one position to the right:

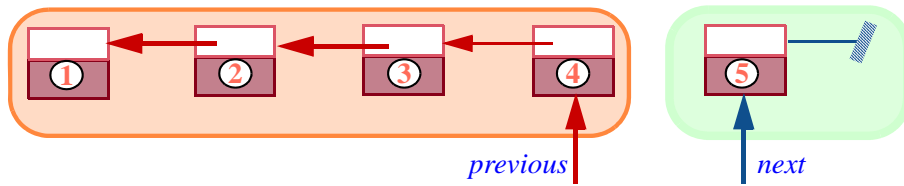

To achieve this change, the loop body will perform a short pointer ballet, which will be detailed below. Note in particular that it must change the *right* field of the first item to the right of the border (in the example, the one with value 4).

Repeating this process, we will eventually reach a state where *next* is void (null) and *previous* points to the last element of the original list, giving us the desired result if we then set *first* to *previous*. The process is easy to initialize: just set *previous* to void and *next* to *first*.

## 2.2 Algorithm text

All common forms of linked list reversal use the scheme just described, with small variations. We will work with the following form:

```
reverse
        -- Rearrange cells into the reverse of their original order.
    local
        previous, next, temp: detachable LINKABLE
    do
        from
            previous := Void ; next := first
        until
            next = Void
        loop
            temp := previous              -- i1
            previous := next              -- i2
            next := next.right            -- i3
            previous.set_right (temp)     -- i4
        end
        first := previous
    end
```

The procedure *set_right* sets the *right* field of its target to the value of its argument. A Java or C# programmer might write the call *previous*.*set_right* (*temp*) as a remote assignment *previous*.*right* := *temp*, but we restrict ourselves to a proper form of O-O programming which rules out such violations of information hiding: the only way to set a field of another object is through a setter procedure such as *set_right*. The **detachable** declaration marks variables whose value might be void [14].

Some simplifications are possible: the initial assignment of **Void** to *previous* is not necessary thanks to default initialization rules; we can get rid of the variable *previous* altogether, and of the final assignment to *first*, by working directly with *first*. We omit these simplifications in the interest of clarity.

For ease of reference in the proof, the four instructions of the loop body have been given names, *i1* to *i4*.

## 2.3 Specification

The first step in verifying software is to specify what needs to be verified. Proper notations are essential: concise, clear, and applicable to a wide class of problems. We need to equip the routine *reverse* with a postcondition stating the property illustrated informally in the preceding figures: that the original list is the concatenation of the list starting at *previous*, reversed, and the list starting at *next*. We express this postcondition as:

$$\textit{first.}{\textstyle\int}\textit{right} \,=\, -\, \textbf{old}\ {\textstyle\int}\textit{first.}{\textstyle\int}\textit{right} \hspace{3em} /1/$$

The expression **old** *e* denotes, as usual, the value of *e* on entry to the routine. If *s* is a sequence (a mathematical object, not a list from programming), $-s$ is the reverse sequence. The "integral" operator $\int$ is a new notation: starting from the current object, $\int b$ denotes the sequence containing that object, then the objects attached to *b*, *b.b*, *b.b.b* and so on, for as long as it makes sense (and stopping at any cycle, although here we are dealing with acyclic structures). The sequence $p.\int b$ similarly contains the objects attached to *p*, *p.b*, *p.b.b* and so on. The integral notation allows us to express the goal of the routine as /1/.

It similarly enables us to express the fundamental invariant property of the loop algorithm. Considered the typical intermediate step, which was illustrated as follows:

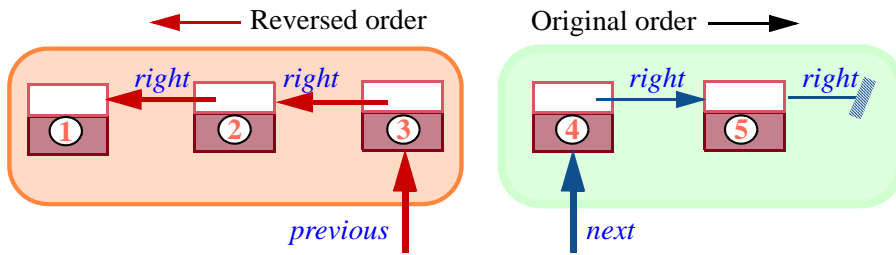

We may express the invariant property that this figure represents as:

$$-\, \textit{previous.}{\textstyle\int}\textit{right}\, +\, \textit{next.}{\textstyle\int}\textit{right}\, =\, \textbf{old}\ {\textstyle\int}\textit{first.}{\textstyle\int}\textit{right} \hspace{3em} /2/$$

where "+" denotes sequence concatenation. Proving this property to be a loop invariant is the key step of a proof of the program. Once we establish this result, it remains only to prove that the invariant is ensured by the initialization (*previous* := **Void** ; *next* := *first*), and that when combined with the loop exit condition (*next* = **Void**) it yields the desired postcondition /1/. Both of these properties are obvious and any good proof machinery will discharge them easily; so the rest of this discussion limits itself to proving that the loop body (when *next* /= **Void**) preserves /2/ and that the loop terminates.

We can in fact simplify the problem further since compositional logic generalizes the notion of loop invariant from boolean expressions to expressions of arbitrary type. We say that an expression *e* is an invariant of a loop simply to mean that an execution of the loop body, performed when the loop exit condition does not hold, preserves its value. (In addition, an invariant of boolean type must have value true after the initialization, and hence will remain true, in keeping with the semantics of traditional invariants.) With this convention the property to prove is that the following expression, which we call *INV*

$$- previous.\!\int\! right + next.\!\int\! right \qquad \text{/INV/}$$

is an invariant of the loop; it no longer needs the **old** operator. (In Eiffel the loop itself would be written

```
from … until … invariant           -- Other clauses as above
     – previous.∫right + next.∫right
loop … end
```

assuming a suitable extension of the language to accept arbitrary expression types in **invariant** clauses.)

A point of notation: in /1/, /2/ and all later assertions involving sequences, the "=" symbol represents mathematical equality, here between two sequences. In an O-O programming language, such assertions will have to use the notation for object equality ("~" in Eiffel, where "=", applied to references, represents reference equality).

## 2.4 Proof approach

In the framework of compositional logic, the property expressing that *INV* is an invariant is

$$(b \,;\, INV) \;=\; INV \qquad \text{-- Where } b \text{ is the loop body: } i1 \,;\, i2 \,;\, i3 \,;\, i4$$

(under the assumption that the exit condition *next* = **Void** does not hold). The notation *i* ; *e*, for an instruction *i* and an expression *e*, denotes the value of *e* after execution of *i*, stated as an expression in the state preceding that execution. Note that the semicolon is also used in its traditional role as separator of sequentially executed instructions, as in *i1* ; *i2*; the two uses reflect, as we will see, the same mathematical operator. If an expression is present, as *INV* here, it must be the last element. (We may think of programming languages such as Algol W and C where a block may end with an expression, following a sequence of instructions, and then evaluates to the value of the expression after execution of the instructions.)

*INV* is the sum (concatenation) – *previous*.∫*right* + *next*.∫*right*. Since the semicolon distributes over "+" as over most operators, the proof that *b* ; *INV* = *INV* can be split into three parts:

- Computing *b* ; *previous*.∫*right* (section 2.5); call the result *bp*.
- Computing *b* ; *next*.∫*right* (section 2.6); call the result *bn*.
- Computing –*bp* + *bn* and showing that it is equal to *INV* (section 2.7).

In addition, section 2.8 will prove loop termination.

The semicolon is right-associative: (*i* ; *j*) ; *e* is *i* ; (*j* ; *e*). As a consequence, since *b* is *i1* ; *i2* ; *i3* ; *i4*, the computation of *b* ; *e4* (where *e4* is *previous*.∫*right* in 2.5 and *next*.∫*right* in 2.6) will proceed as the computation of *e3* = *i4* ; *e4*, then of *e2* = *i3* ; *e3*, then of *e1* = *i2* ; *e2*, then of the result as *i1* ; *e1*. The basic form *i* ; *e* of compositional logic leads to this backward order, recalling how the Hoare assignment axiom leads to backward reasoning.

For ease of reference here is the loop body *b* again:

```
temp := previous              -- i1
previous := next              -- i2
next := next.right            -- i3
previous.set_right (temp)     -- i4
```

One of the attractions of the style of proofs presented in this work is that it closely matches the intuitive semantics of object-oriented programs and the way programmers think about their execution. To take advantage of this property, the reader may find it useful to relate intermediate steps of the proofs to intermediate steps of the computation, as reflected in the illustration of the pointer ballet (on the next page). Going from the bottom up in the figure, the successively computed expressions *e4*, *e3*, *e2* and *e1* correspond to the states S4, S3, S2 and S1.

## 2.5 Handling the *previous* part

We first compute *i4* ; *p* where *p* is *previous*.∫*right* and *i4* is a call to a setter procedure: *previous*.*put_right* (*t*). By coincidence this first step of the proof uses one of the most powerful rules to be seen below, ICX /34/, which states that with a setting procedure *set_a* that sets the value of an attribute *a* in the target object *x*, then

| *x*.*set_a* (*c*) ; *x*.∫*a*     =     <*x*> + *c*.∫*a*, |

where <*x*> denotes the sequence consisting of the single element *x*. We get:

| *i4* ; *p*                = <*previous*> + *temp*.∫*right*                            /p3/ |

We can indeed see at the bottom of the figure how *p* from state S4, that is to say the sequence starting at *previous*, corresponds in state S3 to the element <*previous*> followed by the sequence starting at *temp*.

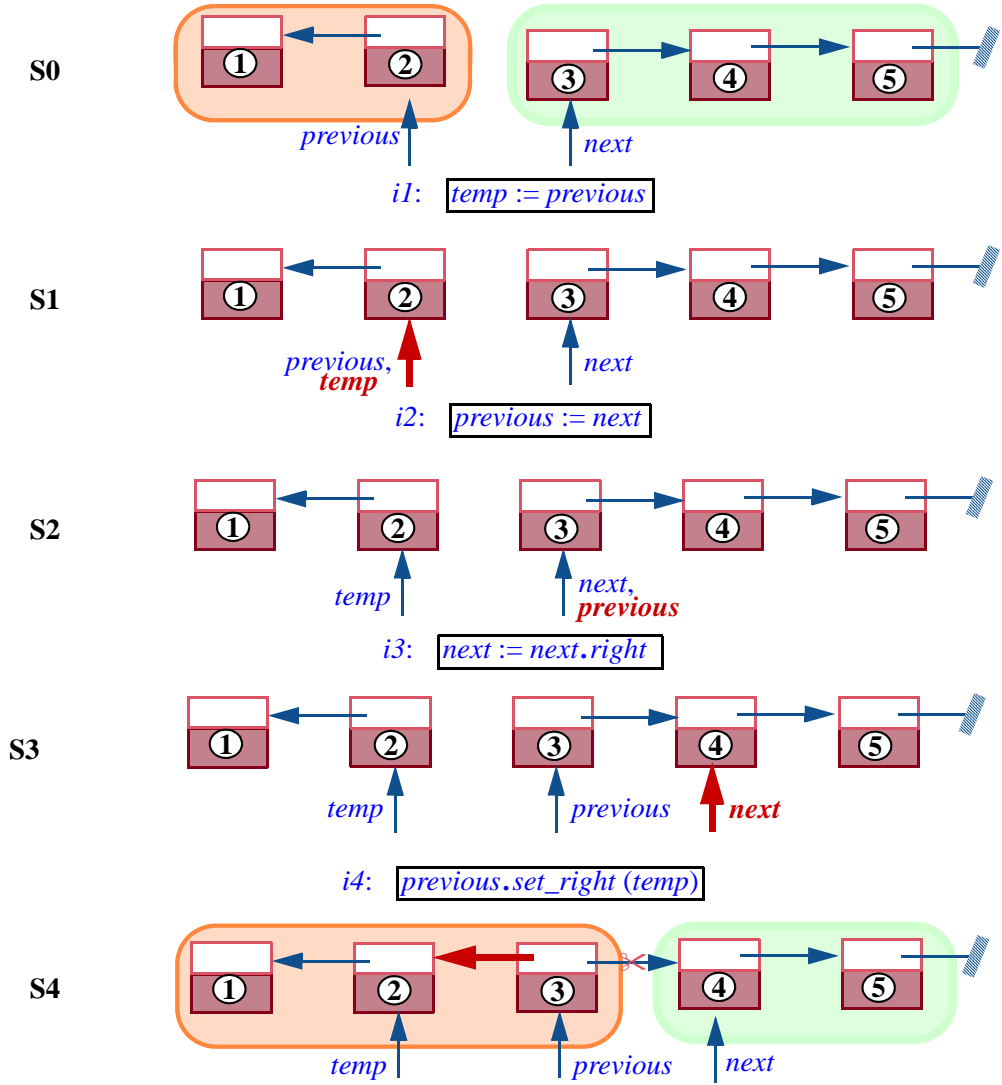

Next we compute *i3* ; *p3* where *i3* is *next* := *next*.*right*. (In this proof the label given for each new step, here *p3*, will denote the value of the expression obtained at that step.) We apply distributivity to compute the effect of the assignment on the two operands of the "+" expression. In both cases, the assignment affects none of the elements in the given expressions, and no cycles are involved; these conditions enable us to apply a theorem seen below, IAY /30/, which indicates that both sides are untouched:

$$i3 \; ; \; p3 \quad = p3 \quad = <previous> + temp.\!\int\! right \qquad /\mathbf{p2}/$$

The figure indeed suggests that the assignment *i3* only affects the "*next*" part of the structure and that the "*previous*" and "*temp*" parts are unchanged between states S2 and S3.

Continuing up the loop body *b*, we compute *i2; p2* where *i2* is the assignment *previous* := *next*. We again consider the two operands separately. The assignment axiom of compositional logic tells us that:

- $((x := e) ; x) = e$; this will be rule AX /5/. It applies here to the first operand since *previous* is the assignment's target.
- For a variable *y* other than *x*, $((x := e) ; y) = y$; this will be rule AY /6/, generalized to expressions as IAY /30/ under conditions of acyclicity satisfied here. It tells us that the assignment has no effect on the second operand.

As a consequence

| | | |
|---|---|---|
| *i2* ; *p2* | = <*next*> + *temp*.∫*right* | /**p1**/ |

We may again perform a visual check on the figure: the sequence that starts with *previous* in state S4 was, in state S1, the concatenation of the *next* element and the sequence starting with *temp*.

In the last proof step, the instruction *i1* is the loop's initial assignment, *temp* := *previous*. Axiom AY /6/ tells us that has no effect on the *next* operand, but axiom IAX /29/ tells us that $(x := y) ; x.\int a$ is (again in the absence of cycles) $y.\int a$. We get as a result the value of *b* ; *p* on entry to the loop:

| | | |
|---|---|---|
| *b* ; (*previous*.∫*right*) | = <*next*> + *previous*.∫*right* | /**bp**/ |

which can again be checked for reasonableness in the figure, by looking at the counterpart in state S0 of the sequence starting with *previous* in state S4. This completes the computation of the effect of *b* on the first operand of our conjectured invariant expression.

## 2.6 Handling the *next* part

We now apply the same process to compute *b* ; *n* where *n* is the second operand, *next*.∫*right*. The reader is invited to follow the intermediate steps in the figure as was done for the first part.

The frame theorem F indicates that the final instruction *i4* of the loop, **call** *previous*.*put_right* (*t*), has no effect on *next*.∫*right*:

| | | |
|---|---|---|
| *i4* ; *n* | = *n* | /**n3**/ |
| | -- where *i4* is "**call** *previous*.*put_right* (*t*)" | |
| | -- and *n* is "*next*.∫*right*". | |

The condition for the rule to be applicable is that *previous* must not be aliased to *next*; the alias calculus yields it here automatically (although we could also establish it through classical techniques).

For *i3*, the assignment *next* := *next*.*right*, theorem IAX /29/ gives the next step:

| | | |
|---|---|---|
| *i3*; *n3* | = *next*.*right*.∫*right* | /**n2**/ |

The initial assignments, *i1* and *i2*, have *previous* and *right* as their respective targets. Rule IAY /30/ tells us that they have no effect on the right side of the above: the next two steps *i2 ; n2* and *i1 ; n1* give the same result as *n2*. As a consequence, we get the final answer for the second operand:

| | |
|---|---|
| *b* ; (*next*.∫*right*)     = *next*.*right*.∫*right* | **/bp/** |

## 2.7 Combining the results

The value we are computing (the value of the conjectured invariant in the initial state of the loop) is – *bp* + *bn*, or, from the preceding computations

$$- (<next> + previous.\!\int right) + next.right.\!\int right$$

Three simple properties of mathematical sequences are that – (*s1* + *s2*) is (– *s2*) + (– *s1*); that concatenation "+" is associative; and that a one-element sequence is its own inverse: – <*a*> = <*a*> for any element *a*. We can use them to simplify the result into

| | |
|---|---|
| – *previous*.∫*right* + <*next*> + *next*.*right*.∫*right* | **/bp1/** |

Theorem SIE /25/, which follows directly from the definition of the integral operator ∫, states that for any attribute *a*

$$\int a \quad\quad\quad = <a> + a.\!\int a$$

indicating that the last two terms in *bp1* combine into *next*.∫*right*, and finally giving us, for the entire expression:

$$- previous.\!\int right + next.\!\int right$$

This is the original expression, completing the proof that the expression is a loop invariant.

## 2.8 Termination

So far the proof has not addressed termination. Informally: since the loop's exit condition is *next* = **Void**, we must make sure that the repeated applications of the loop body finitely reach a void link, thanks in particular to the instruction *i3*: *next* := *next*.*right*. This would not be the case with a cyclic structure; indeed, the routine needs a precondition and should be written as:

```
reverse
            -- Reverse order of the cells.
    require
        first.⊗ right
    … The rest as above …
```

where the property ⊗*a*, for an attribute *a*, states that the structure induced by *a* starting from the current object has no cycle; more generally, *p*.⊗*a* states that the structure induced by *a* starting from *p* has no cycle.

As a consequence of the precondition, the program will maintain the properties *previous*.⊗*right* and *next*.⊗*right*. In other words, the figures showing both the *previous* and *next* lists as acyclic do not lie. This property is proved automatically by application of the alias calculus to the program.

To prove termination formally we need, as usual, a loop variant. If *p*.⊗*a* holds, there is an integer *n*, the "depth of *a* after *p*", written *p*.↓*a*, such that following the *a* links *n* times from *p* leads to an object whose own *a* link is **Void**. In the example *next*.↓*right* is a variant for the loop, guaranteeing termination.

This step completes the example proof, which demonstrates the method developed in this article. We will now review the basis for the properties on which the proof has relied.

## 3  Compositional logic

The first step is to define a proof framework appropriate for reasoning about complex programs. Compositional logic is a variation on the familiar forms of programming language semantics; its main advantage over axiomatic techniques — an advantage of style rather than substance — is that it does not rely on textual substitutions, except in the case of modeling argument passing.

### 3.1  Basics

Compositional logic works with formulae of the following form, for an instruction *i* and an expression *e:*

> *i* ; *e*

denoting the value of *e* after the execution of *i*. For the various kinds of instruction and expression, the rules of compositional logic define *i* ; *e* in terms of expressions evaluated in the state preceding that execution.

As an example, the following axiom applies to any instruction *i* if *c* is a constant:

> *i* ; *c*            = *c*            -- For any instruction *i*    **CONST /3/**

(For ease of reference, all rules appear in shaded boxes and are given both a name and a number.) If we extend this property of constants to arbitrary expressions, we get a generalized version of the concept of "*relative purity*" of an instruction *i* for an assertion *P*, defined in [30] as {*P*} *i* {*P*}: we may say that *i* is relative pure for an expression *e* of any type if (*i* ; *e*) = *e*.

In object-oriented programming, a particularly important constant is **Current** (also called **this** or **self** in various O-O languages), denoting the current object. No construct can ever change the value of **Current**:

| | | | |
|---|---|---|---|
| $i$ ; **Current** | = **Current** | -- For any instruction $i$ | **CUR /4/** |

This rule is our first encounter with the O-O principle of general relativity: as an observer traveling in a spacecraft can change the contents of that vessel but not move to another spacecraft, the execution of an operation on an object can change the contents of that object but not make another object current. (Another analogy is that while you can change some of your own properties you cannot become someone else.)

The CUR property holds of all basic instructions and must be preserved by rules for composite instructions such as calls.

The next two axioms define assignment; for variables $x$ and $y$ and an arbitrary expression $e$:

| | | |
|---|---|---|
| $(x := e)$ ; $x$ | $= e$ | **AX /5/** |
| $(x := e)$ ; $y$ | $= y$ | **AY /6/** |

Here, and elsewhere unless explicitly noted otherwise, different variable names in the axioms, such as $x$ and $y$, denote different variables. (The *values* of the variables could, of course, be equal at run time.)

AX and AY replace the usual assignment axiom of axiomatic semantics. They apply to individual variables rather than arbitrary expressions; to determine the effect of an assignment on a composite expression, we need a distributivity theorem.

## 3.2 Distributivity and associativity

The distributivity theorem

| | | |
|---|---|---|
| $i$ ; $(e \S f)$ | $= (i ; e) \S (i ; f)$ | **DIST /7/** |

is applicable to all ordinary operators § on basic types and references. An example proof using this property and some of the previous ones is:

| | | |
|---|---|---|
| $(x := e)$ ; $(x + 1)$ | $= ((x := e) ; x) + ((x := e) ; 1)$ | -- by DIST /7/ |
| | $= ((x := e) ; x) + 1$ | -- by CONST /3/ |
| | $= e + 1$ | -- by AX /5/ |

In words: the value of $x + 1$ after the assignment $x := e$ is the value of $e + 1$ (computed in the initial state).

An associativity rule applies, where the semicolon is also used in its traditional role as instruction sequencer:

> $(i \,;\, j) \,;\, e \qquad = i \,;\, (j \,;\, e)$                          **ASSOC /8/**
> -- If *e* does not involve **old** (see next)

## 3.3 Rule for "old"

The operator **old** makes it possible to refer to the original value of an expression. The corresponding axiom reflects this property:

> $i \,;\, \textbf{old}\ e \qquad = e$                                            **OLD /9/**

This property holds of all basic instructions *i* and must be preserved by rules for composite instructions such as routine calls.

It must be clear what the scope of *i* is: as stated in the restriction to the ASSOC rule, associativity does not apply if **old** is involved. Compare:

> $(x := 0) \,;\, (x := 1)) \,;\, \textbf{old}\ x \quad = x \qquad$ -- by OLD
> -- but:
> $(x := 0) \,;\, ((x := 1) \,;\, \textbf{old}\ x) \quad = (x := 0) \,;\, x \qquad$ -- by OLD
> $\qquad\qquad\qquad\qquad\qquad\qquad\;\; = 0 \qquad\qquad\;$ -- by AX /5/

As an example of a proof involving **old**, consider the following property:

> $(item := item + 1) \,;\, (item = \textbf{old}\ item + 1)$

which might appear in a class describing a integer counter, whose value is given by *item*. As the expression is to the right of the semicolon is of boolean type, this is the equivalent to proving the Hoare triple {**True**} $(item := item + 1)$ {$item = \textbf{old}\ item + 1$}. Through DIST /7/ applied to the equality operator "=", the property expands to

> $((item := item + 1) \,;\, item) = ((item := item + 1) \,;\, (\textbf{old}\ item + 1))$      **/10/**

The left side of this equality is $item + 1$ by the assignment axiom AX /5/. The right side can be further expanded through DIST to

> $((item := item + 1) \,;\, (\textbf{old}\ item)) + ((item := item + 1) \,;\, 1)$

The first term is *item* by OLD /9/; the second term is 1 by CONST /3/, yielding $item + 1$ for the right side of /10/, and hence establishing /10/.

In comparing this proof with its counterpart in Hoare semantics or weakest-precondition calculus, we may note that it avoids any use of substitution, relying instead on algebraic laws of distributivity and associativity. On the other hand it requires two assignment axioms, AX /5/ and AY /6/, instead of the single axiom of axiomatic semantics.

## 3.4 Calls

Consider a routine $r$. The body of $r$, a sequence of instructions, will be denoted by $\underline{r}$, and the list of formal arguments by $r^{\bullet}$. A call to the routine, with actual arguments $l$, will be written **call** $r\,(l)$. (Modern languages typically do not need the keyword **call**, but we keep it here for clarity.) The compositional logic rule is

$$(\textbf{call}\ r\,(l))\,;\,e \quad = (\underline{r}\,;\,e)\,[r^{\bullet}:l] \qquad\qquad \textbf{UC}\ /11/$$

where $f\,[v:l]$ denotes the expression $f$ with every occurrence of an element in the list of variables replaced by the corresponding element in the list of expressions $l$. This is the only place where compositional logic uses substitution, to represent actual-formal argument association. The rule's name stands for "Unqualified Call"; the version for qualified calls (**call** $x.r\,(l)$) will appear later as QC /21/.

Since rule UC defines the semantics of calls in terms of the semantics of their constituent instructions, it preserves the CUR /4/ and OLD /9/ properties.

## 3.5 Setters

A theorem applies to setter procedures of the form

$$\begin{array}{ll}
set\_a\,(\ldots\,;\,f:T\,;\ldots) & /12/ \\
\quad\text{-- Among other possible actions, set the value of } a \text{ to } f. \\
\textbf{do} \\
\quad anything\_else \\
\quad a := f \\
\textbf{ensure} \\
\quad a = f \qquad\qquad \text{-- There may be other postcondition clauses.} \\
\textbf{end}
\end{array}$$

where $a$ is (in an object-oriented context) an attribute of the enclosing class. We say that a routine with a postcondition clause $a := f$, where $a$ is an attribute and $f$ an argument, is a **setter for** $a$. The theorem is:

$$(\textbf{call}\ r\,(\ldots,\,c,\,\ldots))\,;\,a \quad = c \qquad\qquad \textbf{US}\ /13/$$
$$\text{-- If } r \text{ is a setter for } a$$

("Unqualified Setter" rule). The position of $c$ in the actual argument list is the position of the setting argument, $f$ above, in the formal argument list.

The proof of US immediately from the previous rule UC /11/, and associativity ASSOC /8/ which enables us to ignore whatever *anything_else* does.

It is often important to deduce properties of routines of which we do not have the implementation but only a contract. UC is applicable whenever the routine has the postcondition $a = f$. An informal proof of this property simply notes that the semantics of such a routine does not change if we add the assignment $a := f$ at the end of its body (including if we do this in any order for distinct attributes $a$), so that the previous proof is still applicable.

### 3.6 Mathematical basis

The ";" operator has a simple mathematical meaning. To see it, we start by looking at non-OO (say Pascal-style) programming, then move to an object-oriented context where the idea is the same but the functions' signature more elaborate.

Fundamentally, ";" is a variant of mathematical composition. Let us use the operator "$_m$" to denote the composition of functions or relations; for functions *f* and *g*, their composition $h = f \;_m g$ is such that $h(x) = g(f(x))$. (Frequent mathematical convention lists the functions in the reverse order, but for programming it makes more sense to write them in the order of application.)

Consider first a non-OO framework. $A \rightarrow B$ will denote the set of functions from *A* to *B* where *A* and *B* are arbitrary states. Let *State* be the set of states and *Value* the set of run-time values. An instruction is a function in *State* $\rightarrow$ *State*. (More precisely, it may be a partial function, to account for undefined computations, or a general binary relation, to account for non-deterministic programs; but these cases do not affect the discussion, so we keep "$\rightarrow$" for simplicity.) An expression is a function in *State* $\rightarrow$ *Value*.

The ";" operator in this context is just function composition "$_m$". This also explains why we can apply it both between instructions, as in *i* ; *j*, and between an instruction and an expression, as in the basic formula of computational logic, *i*; *e*. Associativity ASSOC /8/ applies, enabling us to write *i1* ; *i2*; … ; $i_n$ ; *e*, as long as we use an expression only as the last element. The first *n* functions being composed are in *State* $\rightarrow$ *State*, yielding as their composition another function with the same signature; we then compose this result with *e*, of signature *State* $\rightarrow$ *Value*, giving as overall result another *State* $\rightarrow$ *Value* function representing an expression.

In object-oriented programming the signatures are different as a consequence of general relativity: every instruction and expression is relative to a current object, not specified in the class text (and not changeable by it, see CUR /4/). With *Object* representing the set of objects, the signatures are now:

```
Object → State → State        -- For an instruction
Object → State → Value        -- For an expression
```

and the semicolon operator has the following definition, denoting a generalized form of composition where both operands are applied to the same object:

```
i ; f      = λ x: Object | λ s: State | (i (x) m f (x)) (σ)
```

(In other words, $(i \,;\, f)(x)$, applied to a state σ, is the result of applying $f(x)$ to the result of applying $i(x)$ to σ.) As before, *f* can be either an instruction or an expression but the definition is the same, justifying the use of a single operator ";".

### 3.7 Comparison with other semantic description methods

We may assess the level of abstraction of compositional logic against other approaches to defining the semantics of programs and programming languages.

*Denotational* semantics specifies the programming language by explicitly defining, for every kind of instruction *i*, a function in *State* → *State*, and similarly a function in *State* → *Value* for every kind of expression. (For O-O languages, the signatures also involve *Object*.)

*Axiomatic* semantics works at a higher level of abstraction by defining the effect of instructions on boolean properties of the program state (or, for postconditions, of two states). The *weakest-precondition* variant attempts to turn such properties into a calculus whose rules yield the precondition from the construct and the postcondition.

Compositional logic is at a higher level of abstraction than denotational semantics since it does not explicitly manipulate the state, but only talks about the effect of computations on expressions of interest to the programmer. Unlike axiomatic semantics, however, it defines these properties for arbitrary expressions, not just boolean ones.

In the case of boolean expressions, compositional logic reduces to the weakest-precondition calculus: *i* ; *Q* is *i* **wp** *Q* (the weakest precondition of the instruction *i* for the postcondition *Q*).

The correspondence with Hoare-style semantics is similar: the Hoare triple {*P*} *i* {*Q*} expresses that *P* **implies** (*i* ; *Q*), where **implies** is implication between assertions.

## 4 Negative variables: reasoning on object-oriented calls

In the object-oriented style of programming, the basic operation is the "qualified call"

> **call** *x*.*r* (*l*)                                                                                 /14/

which calls the routine *r*, with actual arguments *l*, on the object OX denoted in the current class text by *x*. For the duration of the call, OX will be the current object; the previously current object will become current again upon termination of the call, including any other calls that it may in turn have triggered.

The terms "client object" and "client class" will denote the caller side (the context that issues the above call); "supplier object" and "supplier class" refer to the target object OX and its class:

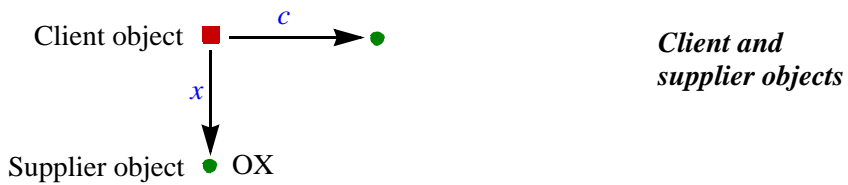

*Client and supplier objects*

The figure (using the precise conventions of "alias diagrams" introduced in [16] for presenting properties of object structures) also shows a field of the client object, corresponding to an attribute *c*, which can be used as an actual argument to the call (part of the list *l*).

The unqualified call rules, such as UC /11/ above, or its equivalent in axiomatic semantics — which tells us that from $\{P\}\ r\ \{Q\}$ we may deduce $\{P\ [r^\bullet: l]\}$ **call** $r\ (l)\ \{Q\ [r^\bullet: l]\}$ —, do not directly apply because they fail to take into account the relativity of expressions in the different contexts of the caller object and the target object. If *f* is a formal argument of *r* (part of $r^\bullet$) and the corresponding actual argument in *l* is *c*, we cannot just substitute *c* for *f* in reasoning about the call, since the name *c* is meaningless for the supplier: it denotes a field of another object, and generally of a different class.

We need, however, to be able to use this field; for example the routine body could perform the instruction

$$y := f.item$$

where *y* is an attribute of the supplier class. In the execution of **call** $x.r\ (c)$, where the formal argument is *f*, we expect this instruction to assign to *y* the value of $c.item$:

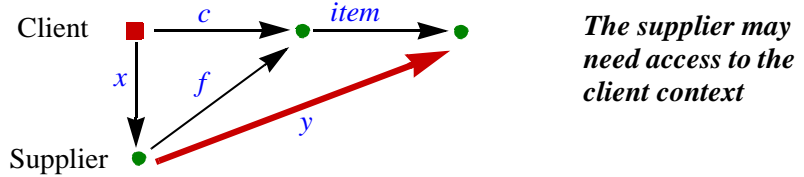

*The supplier may need access to the client context*

Note that *item* itself is a feature of the class of *c*. The thick red arrow in the figure illustrates the intended result of the assignment to *y*. The figure also shows that the formal argument *f* refers, in the supplier's context for this particular call, to the object known in the client's context as *c*.

One way to deal with these changes of context is to assume a preprocessing step in which all unqualified references to features of a class (including attributes, but not formal routine arguments) are prefixed by **Current** (or **this**), then to include in the call rule a substitution of the target, *x* in the example, for all occurrences of **Current**. This is the technique used in [20]. It implies, however, many textual manipulations. We will use instead an algebraic technique based on the notion of negative variable introduced in [16]. The idea is that in a call of target *x* the negated variable *x'*, applicable to the supplier context, denotes a link back to the client object, making it possible in the supplier context to refer to any expression *e* stated in terms of the client context: simply use $x'.e$.

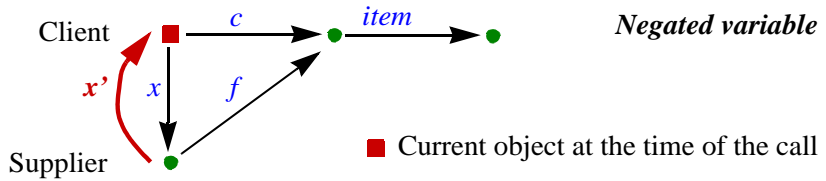

*Negated variable*

■ Current object at the time of the call

For example, passing *c* as the argument in **call** *x*.*r* (*c*) means binding the corresponding formal *f* not to *c* (as in an unqualified call **call** *r* (*c*)) but to *x'*.*c*.

The following rules apply to negative variables and **Current**:

| | | | | |
|---|---|---|---|---|
| *x*.*x'* | = | **Current** | -- For any variable *x* | **NEG1** /15/ |
| *x'*.**old** *x* | = | **Current** | | **NEG2** /16/ |
| **Current**.*e* | = | *e* | -- For any expression *e* | **CUR1** /17/ |
| *e*.**Current** | = | *e* | | **CUR2** /18/ |

(These rules come from [16], with NEG2 adjusted.) The presence of **old** in NEG2 is necessary to account for a "frame" issue: the possibility that a call *x*.*r* (…) might, through a callback, change the value of the *x* field of the current object. Then during the execution of *r*, evaluating *x'*.*x* might lead to the object newly attached to *x*, labeled O' on the next figure, rather than to the call's target OX:

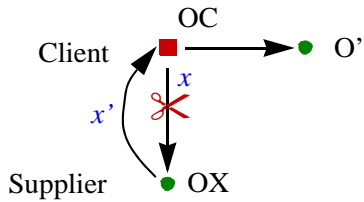

*A callback with side effect can cause x'.x to be no longer* **Current**

Object-oriented languages do permit this behavior, in which a routine call changes the field that served as the call's target; NEG2 handles them. Such schemes complicate verification, however, and break the symmetry between NEG1 and NEG2. It is preferable, as a matter of programming methodology, to avoid them by requiring routines to satisfy the following property:

> **Definition: nonprodigal routine**
> A routine is *nonprodigal* if for any call of target *x* it satisfies the postcondition
> $x'.x =$ **Current**.     **NP** /19/

(The name suggests that the routine preserves its relation with its genitors.) No use of **old** is necessary in NEG1, since the expression *x*.*x'* only makes sense if used from a client during a call of target *x*, and then *x'* always refers back to the client.

The CUR /4/ rule stated that since the current object is given by the context of execution no instruction may ever change the value of **Current**. Similarly, you never get a chance to change the back-link to your client:

| | | | |
|---|---|---|---|
| *i* ; *x'* | = *x'* | -- For any instruction *i* | **BL** /20/ |

(Pursuing the earlier analogy: while you can change other properties of your parents, you cannot become the child of someone else than your parents.)

Negated variables yield a simple semantic description for qualified calls $x.r\ (c)$, the central mechanism of object-oriented computation. Appendix A gives the full semantic rules in both denotational and axiomatic styles. In compositional logic, the rule is

$$(\textbf{call } x.r\ (l))\ ;\ e \qquad = x\ .\ ((\textbf{call } r\ (x'.l))\ ;\ (x'.e)) \qquad \textbf{QC /21/}$$

where "." denotes the dot operator "." distributed over a list (so that $x\ .\ <u, v, \ldots>$ is $<x.u, x.v, \ldots>$). The rule determines how to obtain the effect on $e$ of calling $x.r\ (l)$:

- Transpose the arguments of the original call to the context of the supplier, by prefixing them with "$x'.$". The result of this transposition is **call** $r\ (x'.l)$.
- Find out the effect of this call on the expression $x'.e$, which represents $e$ also transposed to the supplier context. The result is (**call** $r\ (x'.l)$ ; $x'.e$.
- Interpret this result back in the context of the client by prefixing it with "$x.$", giving QC.

This process of transposing the client information to the supplier side then transposing back to the client side directly reflects the unique nature of object-oriented computation with its reliance on the current object. A qualified call makes a new object (the target) current; when the call terminates, the previous current object resumes this role.

Note that if $e$ is **old** $x$, the general rule CUR /4/ governing **Current**, applied to the unqualified call, tells us that **call** $x.r\ (l))$ ; **Current** is **Current.Current** and hence (from CUR1 /17/) **Current**. It follows that the qualified call rule QC also conforms to CUR.

Similarly, (**call** $x.r\ (l))$ ; **old** $x$) = $x$ from QC, NEG2 /16/ and CUR2 /18/. It is not necessarily true, however, that (**call** $x.r\ (l))$ ; $x$) = $x$ because of the frame issue noted above: a callback in the execution of $r$ might modify the client's $x$. field. We may only deduce (**call** $x.r\ (l))$ ; $x$) = $x$ if the routine is nonprodigal as defined above /19/

The QC rule relies on the effect of **call** $r\ (x'.l)$, the unqualified call. That effect is given by the rule for unqualified calls UC /11/, which defines it as the effect of the body after argument substitution. By expanding that earlier rule we get a more detailed version of QC:

$$(\textbf{call } x.r\ (l))\ ;\ e \qquad = x\ .\ ((\underline{r}\ ;\ x'.e)\ [r^\bullet : (x'.l)] \qquad \textbf{QC' /22/}$$

From the qualified call rule (in either form) we get a theorem on qualified calls to setter procedures. As before (section 3.5), we assume that $a$ is an attribute and $set\_a\ (f)$ has the postcondition $a = f$. Then:

$$(x.\textbf{call } set\_a\ (c))\ ;\ ((\textbf{old } x).a) \qquad = c \qquad \textbf{QS /23/}$$

("Qualified Setter" rule, compare with US /13/.) Note that although a setter procedure such as set_a may have several arguments — per the original definition of this notion in section 3.5 — the rest of the discussion ignores, for brevity, any arguments other than one used in a setting role.

The proof of QS is as follows:

$$
\begin{aligned}
(x.\textbf{call}\ set\_a\ (c))\ ;\ ((\textbf{old}\ x).a) &= x.((\textbf{call}\ set\_a\ (x'.\ c)\ ;\ (x'.\textbf{old}\ x.a))) && \text{-- From QC /21/} \\
&= x.((\textbf{call}\ set\_a\ (x'.\ c)\ ;\ (\textbf{Current}.a))) && \text{-- From NEG2 /16/} \\
&= x.((\textbf{call}\ set\_a\ (x'.\ c)\ ;\ a)) && \text{-- From CUR1 /17/} \\
&= x.(x'.\ c) && \text{-- From US /13/} \\
&= \textbf{Current}.c && \text{-- From NEG1 /15/} \\
&= c && \text{-- From CUR1 /17/}
\end{aligned}
$$

Often we may wish a property involving $x$ rather than $\textbf{old}\ x$:

$$(x.\textbf{call}\ set\_a\ (c))\ ;\ x.a\ = c \qquad \text{-- If } r \text{ is nonprodigal} \qquad \textbf{QSN /24/}$$

This property only holds if the routine preserves the link back from its client, as expressed by the "nonprodigal" property NP /19/.

## 5 The alias calculus

The third component of the approach is the alias calculus, developed in an earlier article [16]. For any expressions $e$ and $f$ denoting references, and any program location $pl$, the alias calculus yields the answer to the question: *can the values of $e$ and $f$ ever denote the same object when a program execution is at $pl$*? The theory is (barring any errors in [16]) *sound*, in the sense that if the answer is "no" it provides a guarantee that $e$ and $f$ will always denote different objects — precisely the guarantee we need for the applications discussed here. If the answer is "yes", it could still be the case that $e$ and $f$ never get aliased in practice. In other words, the alias relation that the calculus determines may be an over-approximation of the real aliasings. The possibility of over-approximation comes not from the calculus itself but from the simplification it applies to programming languages: it ignores the conditions in conditionals (defining the aliasings of **if** $c$ **then** $i$ **else** $j$ **end** to be the union of those induced by $i$ and $j$ separately, regardless of $c$) and loops. The over-approximation is generally harmless; when undesired, it can be corrected through the insertion of an assertion $e\ /=\ f$ (expressed in the calculus as the instruction **cut** $e, f$), which needs to be proved, often trivially, through techniques outside of the alias calculus.

The main advantage of the calculus is that its application is automatic. Computing the alias relations induced by a program requires no annotation (except for the occasional **cut**). The calculus yields an algorithm, whose implementation described in [16], although still experimental, covers the entire theory and has been applied to sophisticated examples.

The existence of the alias calculus allows the rest of this discussion to define rules of the form "Property $P$ holds if $e$ and $f$ can never be aliased at the given program point". Such rules are sound — they cannot lead us wrongly to deduce that $P$ holds if it does not — since the calculus is sound.

To express that at a particular program point the expressions *e* and *f*, of reference types, can never be aliased, we will write $e \not\equiv f$. (In Eiffel the usual inequality notation $e \neq f$ suffices, since when applied to references it denotes reference inequality.) This notation has two useful generalizations:

- If *S1* and *S2* are sets or sequences of expressions, $S1 \not\equiv S2$ states that $e \not\equiv f$ for every *e* in *S1* and every *f* in *S2*.
- We may also use $\vec{e} \not\equiv S1$ and $\overrightarrow{e - \{a, b, \ldots\}} \not\equiv S1$ where $\vec{e}$ denotes the set of objects reachable from EO (the object denoted by *e*) by following reference fields any number of times, and $\overrightarrow{e - \{a, b, \ldots\}}$ denotes its subset obtained by starting from fields of EO other than *a, b, …*

These notations are useful to reason about programs, but programmers do not need to know them as they will not appear in assertions or other program elements.

# 6  Reasoning on data structures

It remains to define appropriate concepts and notations to express properties of the kind of object structures, often complex, which routinely arise in object-oriented programming but still defy the reasoning techniques of the usual approaches to program verification.

## 6.1 Background: model-based specifications

One of the reasons for the difficulties experienced by traditional approaches may be that they usually fail to equip themselves with the right abstractions. Typically, they work with elementary values and individual objects. To reason effectively about lists, trees and other sophisticated data structures, we need higher-level abstractions, such as sequences, and we must relate them to the program text; for example, we must be able to refer to the sequence of objects obtained by repeatedly following, from a given object, the successive references of a given type.

The notations defined below, in particular the integral operator, address this requireement. They follow the idea of **model-based specification**, pursued by the author and colleagues [23] [28] but already present in approaches such as JML [11]. This specification method defines the effect of programs in terms of high-level abstractions, representing mathematical concepts (sets, sequences, relations and so on) but closely integrated into the program text and expressed in the host O-O programming language.

In devising these abstractions, we retain one of the key practical properties of the Design by Contract specification method, its support for verification of both the static (proofs) and dynamic (tests) kind, by making sure that contract elements (assertions) not only have a clear mathematical specification but can also be evaluated, under the control of compiler options [6], during program execution.

## 6.2 Context

We assume a statically typed object-oriented language, so that any pointer expression $x \cdot y$ can be considered type-wise valid: there is an attribute of name *x* in the current class, of some type *T*, and in the class defining *T* there is an attribute of name *y*.

References can be "void" (or "null"). We do not need to concern ourselves with the problem of "void calls" (or "null-pointer dereferencing"), even in the absence of a mechanism as Eiffel's Void Safety which removes it entirely at compile time [14], since the conventions defined below will ensure that no void reference is used in an unsafe way.

Object-oriented languages allow attributes from different classes to bear the same names; in fact the Eiffel style rules promote the systematic use of standard attribute names such as *item*. When citing attribute names, the present discussion assumes that they have been disambiguated first, so that each represents an attribute of a single class (and its descendants). Another way of stating this assumption is to assume that every attribute name is prefixed by the name of its class, as in *LINKABLE_item* and *LINKED_LIST_item*.

## 6.3 Paths

The first notion we need (already implicitly used in earlier discussions, with expressions such as $x'.x.c$) is that of a path. A path is a sequence of zero or more attribute names. If the path contains more than one attribute we separate them by periods, as in $a.b.c$.

We may without risk of confusion apply the dot operator to paths (such as $p$ and $q$) as well as attributes (such as $a$, $b$, $c$, $d$, $e$), combining them freely as in $a.p$, $p.a$ and $p.q$, with associativity. For example if $p$ is $a.b.c$ and $q$ is $d.e$, then $p.q$ is $a.b.c.d.e$. This associativity was used in the proof of QS /23/, when it obtained $x.(x'.c)$ and treated it as $(x.x').c$.

A path always denotes an *object*, defined (as the relativistic nature of object-oriented programming requires) in relation to the current object. Informally, we obtain the object denoted by a path $p$ by starting from the current object and following, for as long as possible, the references given by the fields corresponding to the elements of $p$, as in this example:

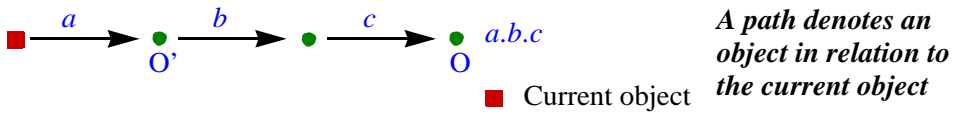

*A path denotes an object in relation to the current object*

"As long as possible" means that the process stops if it encounters a void field, so if the $c$ link in the above figure were void the value of $a.b.c$ would be the same as that of $a.b$. This convention of stopping at void links simplifies the discussion considerably; that it obviously does not reflect the semantics of **Void** or **null** in O-O languages does not matter, since the problem of void safety is not in the scope of the present discussion and should be addressed through separate techniques, such as the void safety framework presented in [14].

To avoid any ambiguity we may define precisely the object O associated with a path $p$:

- If $p$ is empty, O is the current object. In the next two cases, let $a$ (an attribute) be the first element and $q$ the remainder of $p$ (i.e. $p = a.q$, unless $q$ is void in which case $p = <a>$).
- If the $a$ link from the current object is void, O is also the current object.
- Otherwise, let O' be the object to which the $a$ field of the current object is attached. Then $O$ is the object associated (recursively) with $q$ if O' is used as current object.

As this definition unambiguously associates an object with every path, the rest of the discussion often allows itself to talk about "the object $p$" where $p$ is a path.

The *length* $|p|$ of a path $p$ is the number of attributes in its definition. The empty path has length 0 and $a.b.c$ has length 3. In the absence of void links, the length is the number of objects, other than the current object, involved in the path.

If $a$ is an attribute, $a^0$ denotes the empty path, $a^1$ the path $<a>$, and $a^{n+1}$ for $n > 0$ the path $a^n.a$ (which is also $a.a^n$). In line with the general conventions noted above, using this notation assumes proper typing: the type of the attribute $a$ must be the same as the type of the current object, or conform to it.

The notation $\otimes a$ expresses that $a$ is **acyclic**, in the sense that from any current object the sequence $a^n$, for all $n \geq 0$, is acyclic. This property is defined as $a^n \not\equiv$ **Current** for all $n$ — meaning, from the definition of "$\not\equiv$" in section 5, that $a^n$ can never become aliased to the current object (and hence, if the property is satisfied for all possible current objects, that there are no other cycles in the sequence either). One of the principal contributions of the alias calculus to the Calculus of Object Programs is that it tells us, through an automated procedure, that certain attributes are acyclic.

The notation generalizes to $p.\otimes a$, stating that there are no cycles after $p$ in the sequence $p.a^n$.

One more notation is $p.\downarrow a$, the **depth** of $a$ after $p$, defined as the largest $n$ such that all $p.a^i$, for $0 \leq i \leq n$ are different objects. (For empty $p$, we talk of just "the depth of $a$" and write it $\downarrow a$.) This definition covers two cases; calling $s$ the sequence of objects obtained by starting at $p$ and following $a$ links:

- If $s$ is acyclic ($p.\otimes a$ holds), it must reach a void $a$ link: otherwise it would have to be infinite, but our object structures are finite. Then $p.a^n$ is the first object **O** in the sequence whose $a$ field is void.
- If $s$ is cyclic, then $p.a^n$ is the first object **O** in $s$ whose $a$ link leads to **O** itself or a previous element of $s$ ($p.a^i$ for some $i$ in $0..n$).

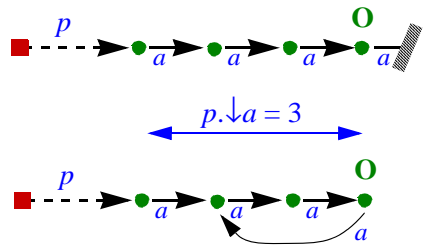

If we know that an attribute is acyclic, the first case applies and we can think of $p.\downarrow a$ as the "distance to **Void** through $a$". As a consequence, $x.\downarrow a$ can serve as the variant for a loop of exit condition $x =$ **Void**, whose body executes $x := x.a$, as in section 2.8 of the example proof.

## 6.4 Integrals

A path denotes a single object. We also need a notation for the *sequence* of objects encountered by repeatedly following the links corresponding to a certain attribute. The integral notation serves that purpose. If $p$ is a path, the notation $p.\int a$ represents the finite sequence of objects $p.a^i$, for all $i$ in $0..n$ where $n$ is $p.\downarrow a$, In other words, it is the sequence of objects that starts with $p$ and continues by following $a$ links, up to the first object in which the $a$ link either is void or leads to an object already in the sequence.

For empty $p$, we write just $\int a$ ("simple integral") denoting a sequence that starts with the current object and continues until the $a$ link would give **Void** or a repetition.

In both cases, the sequence has the following properties:

- It is never empty, since $\int a$ always contains the current object, and $p.\int a$ contains the object associated with $p$ (which always exists as discussed in 6.3).
- It is acyclic by construction.
- It contains objects all of the same type, or of types conforming to a common ancestor type having $a$ as one of its attributes. (This property follows from the assumptions: a typed O-O language, and attribute names that have been disambiguated so that each denotes an attribute of just one class.) In the example, $a$ denotes the same attribute for all objects in the sequence $\int a$ or $p.\int a$.

The notation is inspired by the integrals of classical analysis: as the integral $\int f$ in analysis accumulates the value of the function $f$, our sequence $\int a$ accumulates the values of the attribute $a$. Our integrals can also be compared to regular expressions, but a regular expression denotes a set of sequences, whereas an integral denotes a single sequence.

It would also be possible to define expressions of the form $p.\int a.q$, or even to include several simple integrals, but we do not need such extensions in the present discussion.

Other properties of integrals are:

| | | | |
|---|---|---|---|
| $\int a$ | = | <**Current**> + $a.\int a$ | **SIE /25/** |
| $p.\int a$ | = | $p + p.a.\int a$ | **NIE /26/** |

("Simple Integral Equation" and "Non-simple Integral Equation".) They follow directly from the definitions. Note that the second operand of the "+" is an empty sequence if the $a$ link from (respectively) the current object or $p$ is void.

A general theorem allows us to deduce properties of integrals from properties of paths:

> **Integral theorem**
> Let $p$ be a path and $a$ an attribute; let $f$ a predicate on objects, which can be generalized to a predicate on sequences of objects (which holds if $f$ holds of every element of the sequence). Then we may deduce $f(\int a)$ (resp. $f(p.\int a)$) from any of the following properties:
> 1 • $f(a^n)$ (resp $f(p.a^n)$) for any $n \geq 0$ such that $a^n$ (resp $p.a^n$) is acyclic.
> 2 • $f(q)$ (resp $f(p.q)$) for any path $q$ such that $q$ (resp $p.q$) is acyclic.
> 3 • Either of the previous two without the acyclicity restriction.

To use the theorem, it suffices to show that $f$ holds in the most specific case represented by the first property; in some cases, however, it is just as simple to establish $f$ in the more general cases represented by the second property or even the third.

It is useful to generalize the "may not be aliased" operator "$\not\equiv$" to integrals as follows: $\int a \not\equiv p$, for a path $p$, means that $a^i \not\equiv p$ for all $i$, and similarly for non-simple integrals. To derive such properties, we may as before apply the alias calculus, an automated process.

## 6.5 Compositional semantics of paths and integrals: assignment

It remains to define the effect of instructions on paths and integrals, generalizing the rules defining their effect on simple variables.

There is no simple rule governing the effect of an arbitrary instruction $i$ on a path $p.q$ in the general case. In particular, $i\,;\,(p.q)$ is not necessarily the same as $(i\,;\,p).q$ as illustrated by the following example where $i$ reattaches the $b$ link of O1 from O2 to O3:

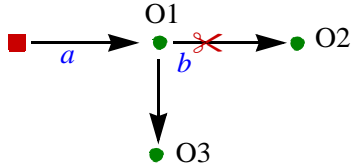

*No associativity for the effect of instructions on paths*

$i\,;\,(a.b)$, is O3, but $(i;\,a)$ is still O1 so $(i\,;\,a).b$ is O2.

We can, however, generalize the assignment rule for single variables (AX /5/ and AY /6/) to paths not involving cycles. The generalized assignment rule is as follows (as usual, $y$ is assumed to denote an attribute other than $x$):

| $(x := e)\,;\,x.p$ | $=$ | $e.p$ | -- If $e.p$ is acyclic | **PAX /27/** |
| $(x := e)\,;\,y.p$ | $=$ | $y.p$ | -- If $y.p$ is acyclic | **PAY /28/** |
|  |  |  | -- See less restrictive conditions below |  |

The basic assignment rules AX and AY (applicable to variables of any type, not just references) are special cases of PAX and PAY for an empty path $p$.

The reason we need an acyclicity restriction is that even though the assignment updates only one field of a single object (the $x$ field of the current object), the $p$ part of the path $e.p$ or $y.p$ could also be affected if it cycles back to that object. The following example shows how a cycle can invalidate PAX. We consider $(x := e)\,;\,x.z.x$ (so that $p$ is $z.x$) under the following circumstances:

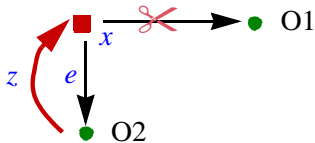

*No associativity for the effect of instructions on paths*

As illustrated, $x$ denotes O1 before the assignment and O2 afterwards. The example assumes that $z$ in O2 points back to the current object. So $(x := e)\,;\,x.z.x$ denotes O2. The value of $e.z.x$, however, is a reference to O1. Here PAX does not hold.

The rules PAX and PAY as stated above require the paths to be acyclic. This condition is stronger than needed since it precludes all cycles, including any that are harmless for the given instruction. A weaker condition suffices: that $x$ (resp. $y$) be *cycle-free for e before p*. This means that no prefix of $p$ is of the form $q.x$ where $e.q$ (resp. $y.q$) may be aliased to **Current**. Acyclic paths are a special case of this condition. Most cases encountered in practice involve paths that are acyclic by construction.

To ascertain acyclicity or cycle-freeness, one may apply the alias calculus.

PAX and PAY have counterparts for integrals:

| $(x := e) ; x.p.\int a$ | $=$ | $e.p.\int a$ | -- If $x$ is cycle-free<br>-- for $e$ before $p$ | **IAX /29/** |
|---|---|---|---|---|
| $(x := e) ; y.p.\int a$ | $=$ | $y.p.\int a$ | -- If $y$ is cycle-free<br>-- for $e$ before $p$ | **IAY /30/** |

These properties assume that $x$ is not the attribute $a$. They follow from extending PAX and PAY through the integral theorem. Having $x$ (resp. $y$) cycle-free for $e$ before $p$ — for example, acyclic — suffices, since the subsequent elements in the sequence, of the form $e.p.\int a$ (resp $y.p.\int a$) result from following $a$ links and cannot be modified by an assignment to an $x$ field of an object.

For the case in which $x$ and $a$ are the same attribute, the following rules apply:

| $(x := e) ; \int x$ | $=$ | $<\mathbf{Current}> + e.\int x$<br>-- If $e.\int x \not\equiv \mathbf{Current}$ | **IA /31/** |
|---|---|---|---|
| $(x := e) ; x.\int x$ | $=$ | $e.\int x$<br>-- If $e.\int x \not\equiv x$ | **IAP /32/** |

(As a reminder, $p.\int a \not\equiv q$ means that $p.a^i$ cannot be aliased to $q$ for any $i$.) These rules are theorems that follow from PAX; in particular the condition of IA, $e.\int x \not\equiv \mathbf{Current}$, follows directly from the condition in PAX: since $\int x$ is by construction acyclic, the only harmful cycles in $e.\int x$ could arise from $e.x^i$ being aliased to **Current** (and similarly for IAP).

The following counter-example shows that these conditions are indeed necessary:

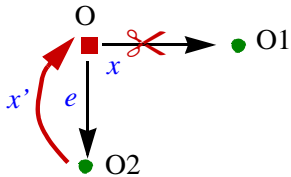

*No associativity for the effect of instructions on paths*

Initially $x$ (the $x$ field of the current object, OS) is attached to O1, $e$ to O2 and the $x$ link of O2 back to O. After the assignment $x := e$, the value of $\int x$ will be the sequence <O, O2>, stopping there because the next $x$ link would cause a cycle. But $<\mathbf{Current}> + e.\int x$ in the initial state was <O, O2, O, O1>. IA does not hold here; indeed its condition is not satisfied since $e.\int x$ was aliased to **Current**.

The six rules just seen enable us to reason about the effect of assignments on paths (for the first two of these rules, PAX and PAY) and integrals. It remains to see the rules defining the effect of calls.

## 6.6 Compositional semantics of paths and integrals: setter calls

The final four rules govern the effect of qualified setter calls **call** $x.set\_a\,(c)$ on paths and integrals. Their application requires some conditions, whose definitions follow; these definitions strive for generality, but it is important to note that any "simple setter" such as *set_right* used in the list reversal example, which simply sets an attribute, trivially satisfies them.

The first two of these rules state the effect of a setter call on a path or sequence starting with the call's target:

| | | |
|---|---|---|
| (**call** $x.set\_a\,(c)$) ; (**old** $x).p$ | $= <x> + c.p$ | PCX /33/ |
| | -- If *set_a* is a setter for *a* | |
| | -- and does not indirectly affect *a*. | |
| (**call** $x.set\_a\,(c)$ ; (**old** $x).\!\int a$ | $= <x> + c.\!\int a$ | ICX /34/ |
| | -- Same condition as previous rule | |

We may apply these rules to *x* rather than **old** *x* if *set_a* is nonprodigal (NP /19/).

The last two rules are "frame conditions" indicating that there is no effect on paths starting with an attribute other than the target:

| | | |
|---|---|---|
| (**call** $x.set\_a\,(c)$) ; $y.p$ | $= y.p$ | PCY /35/ |
| | -- If *set_a* is a setter for *a* | |
| | -- and does not indirectly affect *a*. | |
| (**call** $x.set\_a\,(c)$) ; $(y.\!\int a)$ | $= y.\!\int a$ | ICY /36/ |
| | -- Same condition as previous rule | |

As before, the "P" versions are for paths and the "I" versions for integrals The conditions are defined as follows, for a routine *r* and an attribute *a*:

- Reminder from 3.5: *r* is a *setter* for *a* if it satisfies the postcondition $a = f$, where *f* is one of the routine's arguments.
- The routine is a *simple setter* for *a* if its implementation entirely consists of assignments $f := a$, where *f* is a formal argument of *r*, and possibly of other such assignments of a formal argument to an attribute. The routine *set_right* of list reversal is an example. A simple setter for *a* is a setter for *a*.
- *r directly affects a* if it may change the value of *a*. A setter for *a* affects *a*.
- *r indirectly affects a* if it may change the value of $p.a$ for some non-empty path *p*, or includes a qualified call to a routine that (recursively) affects *a*.

A simple setter for *a* directly affects *a*, and affects no attribute (*a* or another) indirectly.

In the list reversal example, the conditions of the above rules are satisfied since *set_right* is a simple setter for *right*. In addition, a simple setter is nonprodigal, so we can drop the **old** in PCX and ICX. For more general cases, these conditions should be established from the setter's specification:

- The postcondition should express that the routine is a setter.
- It should also limit the scope of changes by expressing that the routine does not indirectly affect the relevant attributes, and possibly that it is nonprodigal; it is preferable, however, to avoid having to state such frame properties explicitly, and rely instead on simple language conventions [18] which imply them.

# 7 Comparison with previous work

The proper handling of references for a verification environment has occupied researchers for a long time. An early paper by Morris [19] defined important steps towards making the problem tractable. Further impetus to research on the topic was spurred by a paper by Hoare and He at ECOOP 99 [9], which took an object-oriented approach. None of the techniques proposed until recently, however, was anywhere close to allowing practical proofs of programs manipulating realistic object structures.

*Separation logic* [24] has enjoyed considerable attention and achieved verification successes. The basic idea is to allow modular reasoning about the heap thanks to the addition to Hoare logic of the $*$ operator, where $P * Q$ means that $P$ and $Q$ separately hold on disjoint parts of the heap. Bornat [1] has published a proof of list reversal using separation logic using C-like programs that manipulate heap addresses directly, quite far from the style of modern object-oriented programming. In recent years, however, there have been applications of separation logic to object-oriented languages, notably [21], [22] and [30], and the development of a proof system based on separation logic, jStar [4]. The main problem with separation logic is the extensive amount of additional annotation that it requires, expressing properties of the heap that are below the level of abstraction at which object-oriented programmers normally work. The corresponding issues are handled in the Calculus of Object Programs through the alias calculus, whose application is automatic. Because of the over-approximation that follows from ignoring conditional and loop conditions in the alias calculus, the results may not be strong enough to allow the desired proofs, in which case the proof engineer will have to add **cut** instructions (section 5 and reference [16]); these instructions are the counterpart, in the Calculus of Object Programs, to the added annotations of separation logic. Their advantage, however, is that (as consistently suggested by experience so far) there will be far fewer of them, and they will only involve specific disjointness properties needed for a particular proof, rather than a complete specification of the heap's state. For example, establishing the alias properties of the proof of linked list reversal in this article required no **cut** instruction whatsoever. This is a good omen for the ease of applying the approach to other applications.

Another property that sets apart the present work from separation logic is its use of properties of object structures, expressed by paths and, through the integral operator, sequences. In separation logic the basic properties of references apply to a single pair of objects, in the form $x \mapsto y$ expressing that the reference in $x$ points to $y$. One of the assumptions behind the preset work is that proofs, and hence specifications, should rely on concepts at a level of abstraction, corresponding to how programmers normally think about their programs; the high-level specification techniques that we have seen above pursue this goal, part of a general scheme of *model-based specification* [29] [23]. Recent work [30]has started to apply separation logic in connection with such specifications. More generally, it is possible that separation logic and the present Calculus of Object Programs could be applied together; the Calculus might for example benefit from the inclusion of some separation logic assertions when it encounters delicate cases. Conversely, the alias calculus may be able to infer or at least suggest separation logic assertions, relieving programmers from having to invent them from scratch.

Another approach that has provided significant advances in the search for techniques to prove object-oriented programs is *dynamic frames* [10] (see also [29] which applies the ideas to an object-oriented language). The theory of dynamic frames addresses the problem of specifying and verifying, in a modular way and in the presence of references, the properties that an operation will *not* modify. While the method is elegant and theoretically attractive, it again requires a significant annotation effort on the programmer's part, to specify frame properties. While it is legitimate, for software reliability, to require programmers to write down the functional specification of the program, it is harder to justify forcing them to state frame properties, since such properties are accessory to the program's real goals. Another objection is that if the program is decently written many frame properties can be inferred automatically from the program text. In the Calculus of Object Programs, the alias calculus is responsible for performing this automatic inference, avoiding the extra specification effort required by dynamic frames. As was noted for separation logic, manual annotations, in the form of **cut** instructions, will only be required if the proof hits a snag; there should be few such cases.

Also like with separation logic, there may be room for combining dynamic frames with the Calculus of Object Programs, for example by using the alias calculus to infer dynamic frame specifications automatically.

Unlike the previous approaches cited — but like the Calculus of Object Programs — *shape analysis* does not require an extensive annotation effort and is instead intended to be automatic. Its roots go back to a long history of work on compiler optimizations, but more recent references [25], [26], [12] have developed it in new directions for the benefit of program verification. (The first two references cited use as an example the list reversal algorithm in a form very close to the version of the present article.) This recent work uses abstract interpretation [3] to construct a Static Shape Graph (SSG) representing an idealized version of the concrete heap. It can then perform analyses of the SSG and relate them back to the actual store; an example, pursuing the same goal as the alias calculus as used here, is a "may-alias" analysis, but the approach can also be applied to many other properties, including proofs, for which an experimental tool, TVLA [27], has been developed. The tool has been applied to a successful automatic proof of a difficult pointer algorithm, Deutsch-Schorre-Waite binary tree traversal [12]. A practical obstacle to using the method, however, is combinatorial explosion of the size of SSGs, resulting in a 9-hour computation time for the example in [12]. The Calculus of Object Programs does not perform any abstraction step but relies on high-level primitives such as the integral operator to capture relevant properties of object structures and reason directly on them in the standard framework of Hoare semantics. It could benefit from the insights of shape analysis; in particular, [26] uses a number of predicates describing high-level properties of object structures: reachability, reachability-from-$x$, sharing, cyclicity, reverse cyclicity. Integrating some of them into the calculus of object structures, in addition to paths and integrals, might increase the expressiveness of the calculus and facilitate proofs.

# 8 Conclusion

The work presented here suffers from several limitations:

- While the alias calculus has been implemented, the rest of the approach has not. It has been designed for integration into an automated proof environment, which should progress quickly.
- The techniques do not yet address inheritance. The main step in adding inheritance is to handle calls to routines that may have several redeclarations in descendant classes. The rules of the Calculus have been defined in reference to specifications of routines — more precisely, their postconditions — rather than their implementation; since these specifications are binding on routine redeclarations through the principles of Design by Contract [13] [5], which limit changes to precondition weakening and postcondition strengthening, their application in the presence of inheritance appears to be a natural extension.
- While the rules should be applicable in a modular way, no particular attention has been devoted to this point as yet.
- The list reversal example is the most significant covered so far. Many more should be tried, involving a variety of data structures.
- The Calculus of Object Structures may need some generalization, for example with a disjunction operator to allow path sets of the form $root.\int(left\ |\ right)$ in a tree example. It has so far been kept as simple as possible.
- On the theoretical side, a proof of soundness is needed to justify the rules of this paper.

All these problems will have to be addressed. I believe, however, that in its present state the Calculus of Object Programs holds the promise of a comprehensive approach to proving full functional correctness of object-oriented programs involving possibly complex run-time object structures. The approach should live up to the claims made on its behalf through the preceding discussion:

- It closely fits the way programmers using modern object-oriented programming languages devise their programs and reason about them.
- The annotations it requires — as any approach addressing functional correctness must — are minimal (alias properties, in particular, are for the most part computed automatically); they express abstract properties of O-O structures, meaningful to the programmer, not low-level descriptions of the makeup of the heap. In fact the notions of heap and stack do not appear, as they are inappropriate at the level of reasoning suitable for modern programming.
- The notations for expressing correctness properties are a small extension to usual Design by Contract mechanisms and remain amenable to run-time evaluation; the approach thereby retains support for both of the dual forms of verification: static (proofs) and dynamic (tests).

The continuing development of the Calculus will endeavor to make these benefits directly available to programers building and verifying object-oriented programs.

## Appendix A: Using negative variables in other semantics

Here is the background for the rules involving negative variables. For the Calculus of Object Programs we only need the rule of compositional logic QC /21/, allowing us to prove properties of programs involving qualified routine calls **call** $x.r\ (l)$, the central mechanism of object-oriented computation. That rule, however, is a consequence of a more fundamental property, giving the denotational definition of qualified calls:

$$\boxed{\textbf{call}\ x.r\ (l) \quad = x \bullet (\textbf{call}\ r\ (x'.l))} \qquad \text{DC /37/}$$

The two sides of the equality are functions in *Object* → *State* → *State*. The rule states that the effect of calling $x.r\ (l)$ is obtained by calling $r$ on arguments transposed to the context of the supplier, as expressed by prefixing them by $x'$, then interpreting the result transposed back to the context of the client, as expressed by prefixing it by $x$. In this result, no occurrences of $x'$ will remain as they go away through the rules on negated variables (NEG1 /15/ to NP /19/).

Some technical notes on this rule:

- The value of **call** $r\ (x'.l)$ is given by the formula for unqualified calls, which states that **call** $r\ (l))\ (\sigma)$ is $(\underline{r}\ (\sigma\ [r^\bullet: l])$. This formula is the basis for the corresponding compositional logic rule UC /11/.
- For a generally applicable form of DC and its unqualified counterpart it is necessary to add to the right side a term that limits the scope of the resulting function to the domain of the original state, getting rid of any temporary associations (affecting for example local variables) that only make sense in the context of the called routine. This restriction is not important for the Calculus of Object Programs.
- DC is an equation rather than a definition, since in the presence of recursion the right-side expression could expand to an expression that includes an occurrence of the left-side expression. Such fixpoint equations are routine in denotational semantics and the theory handles them properly.
- The rule does not require any use of substitution, although it relies on the semantics of the unqualified call **call** $r\ (x'.l)$ which can be defined as $\underline{r}\ [r^\bullet: l]$ (where, following notations introduced earlier, $\underline{r}$ is the semantics of the loop body, $r^\bullet$ denotes the formal arguments, and $e\ [x: y]$ denotes substitution of $y$ for $x$ in $e$).

From this denotational rule we can deduce the axiomatic semantic rule, the object-oriented variant of Hoare's procedure rule [8]:

$$\boxed{\dfrac{\{P\}\ \textbf{call}\ r\ (x'.l)\ \{Q\}}{\{x.P\}\ \textbf{call}\ x.r\ (l)\ \{x.Q\}}} \qquad \text{AC /38/}$$

where $x.e$, for a non-reference expression e (here e is $P$ or $Q$, an assertion. treated as a boolean expression), applies "." distributively, for example $x.(a = b)$ means $x.b = x.a$. In the application of this rule $P$ and $Q$ may contain occurrences of $x'$; for example the rule enables us to deduce {**True**} **call** $x.set\_a\ (c)\ \{x.a = c\}$ from {**True**} **call** $set\_a\ (c)\ \{a = x'.c\}$.

To establish this last property, and more generally the antecedent of any application of AC, we use the ordinary Hoare procedure rule for unqualified calls **call** $r\,(l)$. Expanding this rule (ignoring recursion) in AC gives us a directly applicable version of AC:

$$\frac{\{P\}\ \underline{r}\ (r^{\bullet}\colon x'.l)\ \{Q\}}{\{x.P\}\ \textbf{call}\ x.r\,(l)\ \{x.Q\}} \qquad \text{AC' /39/}$$

Since the denotational rule DC /37/ describes the nature of object-oriented calls at the most fundamental level, we may use it to express properties of such calls in any semantic framework. More generally, let $\Pi$ be a property of program elements, such that the dot operator "**.**" distributes over $\Pi$. Then we may use the general rule

$$\Pi\,(\textbf{call}\ x.r\,(l)) \quad = x\, \blacksquare\, \Pi\,(\textbf{call}\ r\,(x'.l)) \qquad \text{GC /40/}$$

AC, the axiomatic rule, is just one instance of GC. Another instance appears in the alias calculus article [16], which for the various kinds of instructions $i$ and an arbitrary relation $a$ (a set of pairs of expressions that might become aliased to each other) defines $a \gg i$, the alias relation resulting from executing $i$ in a state where the alias relation was $a$. The rule for qualified calls (with "$\blacksquare$" denoting "**.**" distributed over a set of pairs) is

$$a \gg \textbf{call}\ x.r\,(l) \quad = \quad x\, \blacksquare\, ((x'\, \blacksquare\, a) \gg \textbf{call}\ r\,(x'.l))$$

A final example of applying GC is the weakest precondition rule for qualified calls (using $i$ **wp** $Q$ to denote the weakest precondition ensuring that execution of the instruction $i$ will ensure the postcondition $Q$):

$$(\textbf{call}\ x.r\,(l))\ \textbf{wp}\ x.Q \quad = x\, \blacksquare\, ((\textbf{call}\ r\,(x'.l))\ \textbf{wp}\ Q) \qquad \text{GC /41/}$$

The simplicity of these rules appears to confirm the usefulness of negative variables as a tool for reasoning about object-oriented computations.

## Appendix B: References

**34** TOWARDS A CALCULUS OF OBJECT PROGRAMS §8[18] Bertrand Meyer, *If I'm not pure, at least my functions are*, blog entry at bertrandmeyer.com/2011/07/04/if-im-not-pure-at-least-my-functions-are/, 4 July 2011 (intended as a first step to an actual article on language conventions to specify purity and, more generally, frame properties).

[19] Joseph M. Morris: *A General Axiom of Assignment; Assignment and Linked Data Structure*; *A Proof of the Schorr-Waite Algorithm* (three articles), in Theoretical Foundations of Programming Methodology, Proceedings of the 1981 Marktoberdorf Summer School, eds. M. Broy and G. Schmidt, Reidel, 1982, pp, 25-61.

[20] Peter Müller: Modular *Specification and Verification of Object-Oriented Programs*, Springer Verlag, 2002.

[21] Matthew Parkinson and Gavin Bierman: *Separation Logic and Abstraction*, in *POPL '05* (ACM Symposium on Principles of Programming Languages), January 2005, pages 247-258.

[22] Matthew Parkinson and Gavin Bierman: *Separation Logic, Abstraction and Inheritance*, in *POPL '08* (ACM Symposium on Principles of Programming Languages), January 2008, pages 75-86.

[23] Nadia Polikarpova, Carlo Furia and Bertrand Meyer: *Specifying Reusable Components*, in *Verified Software: Theories, Tools, Experiments* (VSTTE ' 10), Edinburgh, UK, 16-19 August 2010, Lecture Notes in Computer Science, Springer Verlag, 2010.

[24] John C. Reynolds: *Separation Logic: A Logic for Shared Mutable Data Structures*, in *Logic in Computer Science*, 17th Annual IEEE Symposium, 2002, pages 55-74.

[25] Mooly Sagiv, Thomas Reps and Reinhard Wilhelm: *Solving shape-analysis problems in languages with destructive updating*, in TOPLAS (*ACM Transactions on Programming Languages and Systems*), vol. 20, no. 1, January 1998, pages 1-50.

[26] Mooly Sagiv, Thomas Reps and Reinhard Wilhelm: *Parametric shape analysis via 3-valued logic*, in *ACM Transactions on Programming Languages and Systems*, vol. 24, no. 3, May 2002, pages 217–298.

[27] Mooly Sagiv et al., TVLA home page, at www.math.tau.ac.il/~tvla/.

[28] Bernd Schoeller, Tobias Widmer and Bertrand Meyer: *Making Specifications Complete Through Models*, in *Architecting Systems with Trustworthy Components*, eds. Ralf Reussner, Judith Stafford and Clemens Szyperski, Lecture Notes in Computer Science, Springer-Verlag, 2006.

[29] Bernd Schoeller: *Making Classes Provable through Contracts, Models and Frames*, PhD thesis, ETH Zurich, 2007, se.inf.ethz.ch/old/people/schoeller/pdfs/schoeller-diss.pdf.

[30] Stephan van Staden, Cristiano Calcagno and Bertrand Meyer: *Verifying Executable Object-Oriented Specifications with Separation Logic*, in *ECOOP 2010*, 24th European Conference on Object-Oriented Programming, Maribor (Slovenia), 21-25 June 2010, Lecture Notes in Computer Science, Springer-Verlag, 2010.

## Acknowledgments

The software verification work of the Chair of Software Engineering at ETH Zurich has been supported by a generous grant from the Hasler Foundation as part of the MANCOM program, by several grants from the Swiss National Science Foundation (FNS/SNF), and by two internal research grants (TH-Gesuch, now ETHIIRA) from ETH Zurich. While not directly performed in response to any of these grants, the work reported here would not have been possible without them.

Part of the work was carried out in the Software Engineering Laboratory of ITMO State University in Saint Petersburg, which also provides an excellent environment.